# New Method for Accurate Parameter Estimation of Induction Motors Based on Artificial Bee Colony Algorithm

Mohammad Jamadi
Zanjan, Iran
Email: jamadi.mohammad@yahoo.com

Farshad Merrikh-Bayat
University of Zanjan/Department of Electrical and Computer Engineering, Zanjan, Iran
Email: f.bayat@znu.ac.ir

*Abstract*—**This paper proposes an effective method for estimating the parameters of double-cage induction motors by using Artificial Bee Colony (ABC) algorithm. For this purpose the unknown parameters in the electrical model of asynchronous machine are calculated such that the sum of the square of differences between full load torques, starting torques, maximum torques, starting currents, full load currents, and nominal power factors obtained from model and provided by manufacturer is minimized. In order to confirm the efficiency of the proposed method the results are also compared with those achieved by using GA, PSO, and PAMP. The simulations show that in the problem under consideration ABC converges considerably faster than other algorithms and the results are as accurate as PAMP.**

## I. Introduction

Induction motors are widely used in industrial applications according to the simplicity of design, high efficiency and low cost. Previously, DC motors were often preferred when exact control was needed, but currently according to the advances in power electronics, effective control of induction motors is also possible and economic. This fact has increased the attentions to this kind of machines in recent years. This increased attention also justifies the importance of accurate modeling of induction motors.

So far various methods have been proposed by researchers for modeling and parameters estimation of induction motors. The classical approach is to extract the parameters of model from no-load and locked-rotor tests [1]. The main drawback of this method is the need for experimental studies, which are sometimes difficult to perform. Some other methods can also be found in the literature for this purpose. For example, numerical methods, which are often tedious, are also used for estimating the parameters of double-cage induction motors [2]. A classical least-square based method for parameter estimation of induction motors can be found in [3-4], and a study on parameter estimation and simulation of double-cage induction motors based on the data provided by manufacturer is presented in [5-6]. Meta-heuristic optimization algorithms such as Genetic Algorithm (GA) [7], Artificial Neural Networks (ANNs) [8], Particle Swarm Optimization (PSO) [9], Artificial Immune System (AIS) [10], Bacterial Foraging Optimization (BFO) [11], shuffled frog-leaping algorithm [12], and Differential Evolution (DE) algorithm [13] are also used for parameter estimation of induction motors. Application of shuffled frog-leaping algorithm for parameter estimation of double-cage induction motors is studied in [14].

It should be noted that the above mentioned methods suffer from drawbacks such as slow convergence, low accuracy and large number of tuning parameters in some cases. Moreover, double-cage induction motors have received less attention in previous works. The aim of this paper is to study the application of Artificial Bee Colony (ABC) algorithm [15] for parameter estimation and modeling of double-cage induction motors, and compare the results with competing methods. Nowadays, this algorithm is used in a wide variety of real-world applications (see for example [16]-[18] and the references therein for more information). The method used in this paper for parameter estimation of double-cage induction motors works based on the minimization of a suitably chosen cost function. A comparison with three other methods (i.e., GA, PSO and PAMP) is also presented to confirm the efficiency of the proposed method. These results show that ABC acts very fast compared to other algorithms and its estimations are fairly close to those achieved by using PAMP.

The reset of this paper is organized as the following. Section II is devoted to the statement of problem. The cost function under consideration is also introduced in this section. A brief review of ABC algorithm is presented in Section III. Simulation results are presented in Section IV, and finally Section V concludes the paper.

## II. PROBLEM STATEMENT

The method proposed in this paper for parameter estimation of double-cage induction motors needs some data of the motor under consideration. These data (including the full load torque, starting torque, maximum torque, starting current, full load current, and nominal power factor) are often provided by the manufacturer and assumed to be known (all similar papers make use of the same assumptions).

Fig. 1 shows the electrical model of the double-cage induction motor. Our aim here is to find the unknown parameters $(X_{2d}, X_{1d}, R_2, R_1, X_m, X_{sd}, R_s)$ in this figure such that the resulted model behaves as close as possible to the corresponding real-world motor. The standard method for this purpose is to calculate the value of these unknown parameters such that the sum of the square of differences between the information extracted from model and the manufacturer data is minimized (of course, this minimization is subjected to the constraints of problem). Hence, it is natural to define the cost function as the following:

$$F = F_1^2 + F_2^2 + F_3^2 + F_4^2 + F_5^2 + F_6^2, \tag{1}$$

Where:

$$F_1 = \frac{T_{fl,cal} - T_{fl,mf}}{T_{fl,mf}}, \quad F_2 = \frac{T_{st,cal} - T_{st,mf}}{T_{st,mf}}, \quad F_3 = \frac{T_{Max,cal} - T_{Max,mf}}{T_{Max,mf}}, \tag{2}$$

$$F_4 = \frac{PF_{fl,cal} - PF_{fl,mf}}{PF_{fl,mf}}, \quad F_5 = \frac{I_{st,cal} - I_{st,mf}}{I_{st,mf}}, \quad F_6 = \frac{I_{fl,cal} - I_{fl,mf}}{I_{fl,mf}}. \tag{3}$$

In the above equations the *fl*, *st*, *Max*, *mf*, and *cal* indices refer to the full load, starting, maximum, manufacturer data, and calculated variable respectively, and the variables *T*, *PF*, and *I* stand for the torque, power factor, and current, respectively. According to the above discussion, in the following it is assumed that the value of variables with *mf* index is provided by the manufacturer and the value of variables with *cal* index is calculated from the model. In order to calculate the torques appear in (2) first we calculate the currents from the following equations:

$$I(s) = \frac{V_{ph}}{R_s + jX_{sd} + Z_p(s)}, \quad I_1(s) = \frac{Z_p(s)I(s)}{R_1/s + jX_{1d}}, \quad I_2(s) = \frac{Z_p(s)I(s)}{R_2/s + jX_{2d}}, \tag{4}$$

Where, according to Fig. 1, $Z_p(s)$ is obtained as the following

$$Z_p(s) = \frac{1}{\dfrac{1}{jX_m} + \dfrac{1}{R_1/s + jX_{1d}} + \dfrac{1}{R_2/s + jX_{2d}}}. \tag{5}$$

After calculation of currents, the torques in (2) are calculated from the following equations:

$$T(s) = \frac{3p}{\omega_s}\left([I_1(s)]^2 \frac{R_1}{s} + [I_2(s)]^2 \frac{R_2}{s}\right), \tag{6}$$

and $S_m$ (used for calculation of $T_{Max,cal}$) is obtained through the following equation

$$\frac{dT(S_m)}{dS} = 0. \tag{7}$$

To sum up, the unknown parameters in the model of double-cage induction motor (as shown in Fig. 1) can be estimated by minimizing the cost function given in (1). However, the following inequality constraints should also be satisfied during this optimization:

$$R_s, X_{sd}, X_m, R_1, R_2, X_{1d}, X_{2d} > 0, \tag{8}$$

$$R_2 > R_1, \tag{9}$$

$$X_{1d} > X_{2d}, \qquad (10)$$

$$\left| \frac{T_{Max,cal} - T_{Max,mf}}{T_{Max,mf}} \right| \leq 0.2 . \qquad (11)$$

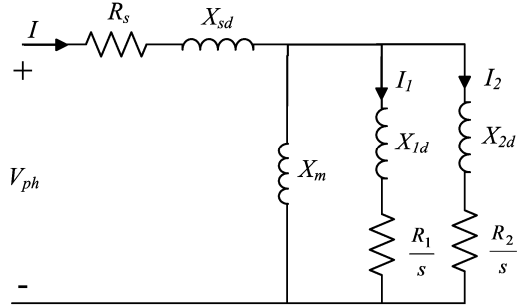

Figure 1.　　Electrical model of the double-cage induction motor

III. Brief Review of Artificial Bee Colony Algorithm

Artificial bee colony algorithm introduced in 2005 by Karaboga [15] and belongs to the family of nature-inspired meta-heuristic optimization algorithms. This algorithm is inspired from the behavior of honey bees in nature and provides us with a powerful tool for solving complex optimization problems. In the ABC algorithm artificial bees in the colony are divided into three parts: employed bees, onlooker bees, and scout bees. Employed bees discover the food sources, bring the food to hive and share its location with other bees. Onlooker bees stay in the hive and decide to follow the employed bees based on the quality of the food sources they have discovered. Scout bees randomly search the outdoor (independent of employed bees) to find (probably better) unseen food sources. In ABC algorithm the location of each food source identifies a point in the domain of problem (i.e., a potential solution) and points with smaller value for cost function are assumed to be better food sources (better solutions).

Mathematically, in the first step of algorithm the solution vectors are selected randomly from the domain of problem. For this purpose, position of the $n$th artificial bee ($n = 1, 2, \ldots, SN$) is considered as the following:

$$X_n = [x_{n1}, x_{n2}, \ldots, x_{nm}], \qquad (12)$$

where $SN$ and $m$ stand for the number of bees and number of variables, respectively, and $x_{n1}, x_{n2}, \ldots, x_{nm}$ are random numbers selected from the domain of definition of problem. At the next step, employed bees search around the food sources $X_n$ (i.e., the previous solutions in their memory) to find the potentially better sources $V_n = [v_{n1}, v_{n2}, \ldots, v_{nm}]$, where the components of $X_n$ and $V_n$ are related through the following equation:

$$v_{ni} = x_{ni} + \varphi_{ni}(x_{ni} - x_{ki}) . \qquad (13)$$

In the above equation $k$ ($k \neq n$) is a randomly selected integer in the range $[1, SN]$ and $\varphi_{ni}$ is a random number (with uniform distribution) selected from $[-1, 1]$. After calculating the location of new sources from (13), the fitness of each new source is calculated from the following equation:

$$fit(X_n) = \begin{cases} \dfrac{1}{1 + f(X_n)} & f(X_n) \geq 0 \\ 1 + \mathrm{abs}(f(X_n)) & f(X_n) < 0 \end{cases}, \qquad (14)$$

where $fit(X_n)$ is the fitness of the source located at $X_n$, and $f(X_n)$ is the value of the ($m$-variable) cost function to be minimized at this point. After calculating the location of new sources and evaluating the fitness of each new source (location), onlooker bees in the hive choose the employed bees of the next iteration based on the quality of their food sources. More precisely, first the probability of choosing the food source located at $X_n$ (to be used in the next iteration for further search around), denoted as $P_n$, is calculated as the following

$$P_n = \frac{fit(X_n)}{\sum_{k=1}^{SN} fit(X_n)}. \tag{15}$$

Then a roulette wheel is used for determining the food sources to be used by employed bees in the next iteration (angle of sectors of the roulette wheel is considered proportional to the probabilities calculated from (15)). Note that at each iteration onlooker bees select exactly *SN* bees by chance, and consequently, some of the employed bees may not be selected at all while some others are selected more than once. In the standard ABC algorithm, one of the employed bees is selected and classified as the scout bee [19] (this definition is slightly modified later [20]). The classification is controlled by a control parameter called ''limit''. In this manner if a solution representing a food source is not improved by a predetermined number of successive trials, then that food source is abandoned by its employed bee and the employed bee associated with that food source becomes a scout, which searches around randomly. The number of trials for releasing a food source is equal to the value of ''limit'', which is an important control parameter of ABC algorithm. In this paper, however, for increasing the accuracy of results we have adopted more that one scout bee (similar to [20]), and moreover, the scout bees follow the best employed bee of colony instead of performing a random search. More precisely, any employed bee that cannot find a better solution (compared to its previous finding) after ten successive iterations is considered as a scout bee.

To sum up, at each iteration first the new locations are calculated from (13), and their qualities and selection probabilities are evaluated through (14) and (15), respectively. Then, some of these locations are selected by onlooker bees and the possible scout bee is determined, and this procedure is repeated until a certain termination condition is fulfilled. It is obvious that by using this procedure low quality food sources are most likely to be abandoned by onlooker bees and, as a result, employed bees tend to search around the locations with higher fitness values. The location with the highest fitness value (taking into account all iterations and all bees) is considered as the final solution (of course, for this purpose the best solution should be memorized at each iteration).

## IV. SIMULATION RESULTS

In this section we use ABC algorithm to solve the constraint optimization problem defined through (1) and (8)-(11), and consequently, find the parameters of the model shown in Fig. 1 (parameters of the algorithm are assumed as $\varphi_{ni} = 0.5$ and $SN = 120$, and the simulation stops after 100 iterations). We will also use PAMP (PAMP is the short for power_AsynchronousMachineParams command in MATLAB, which opens a graphical user interface (GUI) that can be used to estimate the parameters of double-cage asynchronous machines [21]), GA, and PSO for this purpose and compare the performance of resulted models.

Table I shows the manufacturer data of the double-cage induction motor under consideration. Parameters of the model shown in Fig. 1 are estimated by applying four different optimization methods and the results are summarized in Table II. In this table comparing the results of PAMP with the corresponding ones obtained by applying ABC, PSO, and GA reveals the fact that ABC leads to the more accurate results compared to GA and PSO (note that in practice we do not have access to the exact value of parameters, and consequently, we can only compare the results with PAMP, which is believed to be the most accurate method). Each method in Table II leads to a special model for motor in the general form of Fig. 1, which can be used for estimation of other variables such as the maximum torque, nominal torque, starting torque, starting current, nominal current and nominal power factor of motor. These variables are also calculated in four cases separately and the results are presented in Table III. The corresponding data provided by manufacturer is also presented in this table for the sake of comparison. As it can be observed in this table, the ABC algorithm works considerably better than others since it leads to (almost in all items) smaller estimation errors. In fact, GA leads to the less accurate results compared to other methods, and ABC's estimation of some parameters is even better than PAMP's since they are closer to the manufacturer data. Note that in dealing with each algorithm the simulations are performed several times and the best result is used for comparison purposes.

TABLE I.
MANUFACTURER DATA OF A 2200W DOUBLE-CAGE INDUCTION MOTOR [21]

| $T_{St}$(Nm) | $T_{fl}$(Nm) | $T_{Max}$(Nm) | $I_{st}$(A) | $I_{fl}$(A) | $Pf_{fl}$ | $S_{fl}$ | V | f (Hz) |
|---|---|---|---|---|---|---|---|---|
| 43.31 | 12.27 | 47.73 | 66.48 | 8.3 | 0.87 | 0.039 | 208 | 60 |

TABLE II.
PARAMETER ESTIMATION OF DOUBLE-CAGE INDUCTION MOTOR USING DIFFERENT METHODS

|  | ABC | PSO | GA | PAMP |
|---|---|---|---|---|
| $R_1$ | 1.1855 | 1.1855 | 0.919 | 1.183 |
| $X_{sd}$ | 0.1259 | 0.1146 | 0.6973 | 0.1257 |
| $X_m$ | 25.077 | 23.572 | 23.7683 | 25.4211 |
| $X_{1d}$ | 0.1299 | 0.1154 | 0.6543 | 0.1573 |
| $R_{1d}$ | 1.1648 | 1.1852 | 1.0485 | 1.253 |
| $X_{2d}$ | 0.1187 | 0.1145 | 0.3057 | 0.1257 |
| $R_{2d}$ | 1.3641 | 1.4287 | 1.6471 | 1.257 |

TABLE III.
PARAMETER ESTIMATION OF DOUBLE-CAGE INDUCTION MOTOR USING DIFFERENT METHODS

| Parameter | Manufacturer Data | PAMP | PAMP Error (%) | ABC | ABC Error (%) | PSO | PSO Error (%) | GA | GA Error (%) |
|---|---|---|---|---|---|---|---|---|---|
| $T_{st}$ (Nm) | 43.31 | 43.57 | 0.60 | 43.5 | 0.44 | 43.98 | 1.55 | 44.8588 | 3.58 |
| $T_{fl}$ (Nm) | 12.27 | 12.57 | 2.44 | 12.55 | 2.28 | 12.22 | 0.40 | 13.2721 | 8.14 |
| $T_{Max}$ (Nm) | 47.73 | 48.33 | 1.26 | 48.68 | 1.99 | 50.32 | 3.44 | 51.870 | 8.67 |
| $I_{st}$ (A) | 66.48 | 65.8839 | 0.902 | 65.8059 | 1.01 | 65.1745 | 1.96 | 64.2793 | 3.31 |
| $I_{fl}$ (A) | 8.3 | 8.2933 | 0.12 | 8.3196 | 0.23 | 8.3182 | 0.21 | 8.6064 | 3.69 |
| $Pf_{fl}$ | 0.87 | 0.8747 | 0.57 | 0.8716 | 0.18 | 0.8508 | 2.24 | 0.8729 | 0.33 |

In order to make a better comparison, the slip-torque plots of the resulted models are shown in Fig. 2. Three data points provided by manufacturer (corresponding to starting, maximum and nominal torques) are also shown in this figure. It is observed that the curves corresponding to ABC, PSO, and PAMP are fairly matched with the data points provided by manufacturer. Fig. 3 shows the slip-current plots of the resulted four models, where the starting and nominal currents provided by manufacturer are denoted as stars. As it can be observed in this figure the error caused by ABC at $s=1$ is almost the same as PAMP at this point and both of them are more accurate than GA and PSO. The starting current estimated by ABC is also slightly closer to the manufacturer data compared to others. Finally, Fig. 4 shows the minimum value obtained for objective function versus the number of iterations. This figure clearly shows the faster convergence of ABC compared to other methods (ABC reaches its final value after about 10 iterations).

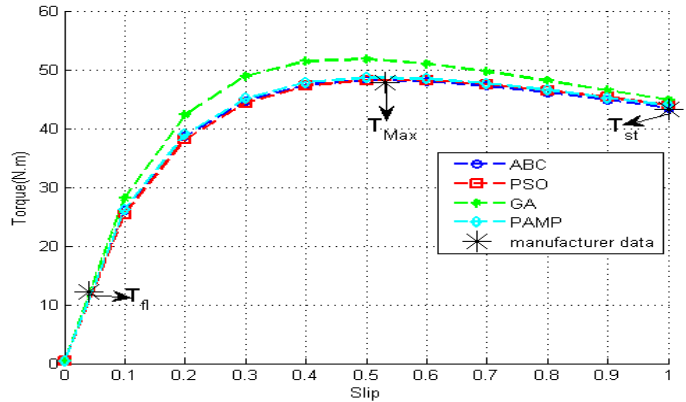

Figure 2. Slip-torque plots of the models obtained by using different optimization algorithms

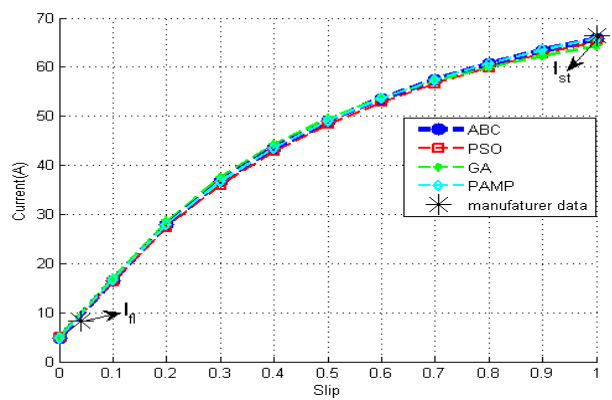

Figure 3. Slip-current plots of the models obtained by using different optimization algorithms

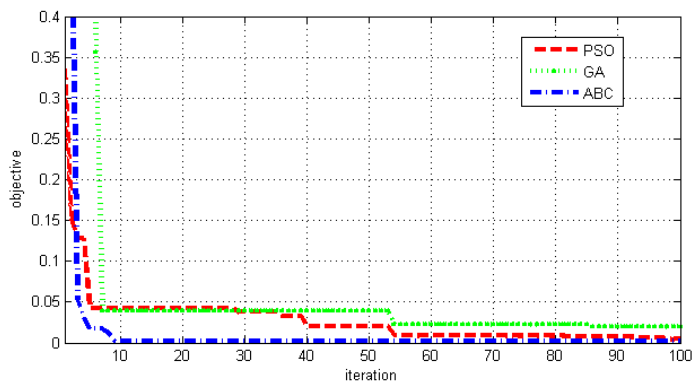

Figure 4. Objective function value versus the number of iterations

## V. Conclusion

In this paper we studied the application of ABC algorithm for estimating the parameters of a 2200 w double-cage induction motor and compared the results with those obtained by using GA, PSO and PAMP. For this purpose, the parameter estimation problem is expressed in terms of a constrained optimization problem in which the unknown parameters of model are calculated such that the sum of the square of differences between starting torques, nominal torques, maximum torques, starting currents, nominal currents, maximum currents and nominal power factors of the model and the manufacturer data is minimized. Simulations show that ABC can estimate the parameters of model as accurate as PAMP and more accurate than GA and PSO. Moreover, ABC converges to the optimal solution considerably faster than GA and PSO. It concludes that ABC is an effective tool for parameter estimation and modeling of double-cage induction motors.

## References

[1] Say, M. G. (1983). Alternating current machines (5th Ed.). Pitman Press.
[2] Pedra, J. & F. Corcoles (2004). Estimation of induction motor double-cage model parameters from manufacturer data. *IEEE Transaction on Energy Conversation.* 19.2, 310-317.
[3] Koubaa, Y. (2004). Recursive identification of induction motor parameters. *Simulation Modelling Practice and Theory.* 12.5, 363-381.
[4] Stephan J., M. Bodson, & J. Chiasson (1994). Real time estimation of induction motor parameters. *IEEE Transaction Inddustry Application.* 30.3, 746-759.
[5] Ansuj, S., F. Shokooh & R. Schinzinger (1989). Parameter estimation for induction machines based on sensitivity analysis. *IEEE Transaction Inddustry Application.* 25-6, 1035-1040.
[6] Lindenmeyer, D., H. W. Dommel, A. Moshref & P. Kundur (2001). An induction motor parameter estimation method. *International Journal of Electrical Power & Energy System.* 23, 251-262.

[7] Alonge, F., F. Dippolito, G. Ferrante & F. M. Raimondi (1998). Parameter identification of induction motor model using genetic algorithms. *IET JOURNALS.* 145.6, 587-593.
[8] Bae, D (1997). Determination of induction motor parameters by using neural network based on FEM results. *IEEE Transaction on Magnetics.* 33, 1924-1927.
[9] Sakthivel, V. P., R. Bhuvaneswari & S. Subramanian (2010). Multi objective parameter estimation of induction motor using particle swarm optimization. *Engineering Applications of Artificial Intelligence.* 23.3, 302-312.
[10] Sakthivel, V. P., R. Bhuvaneswari & S. Subramanian (2010). Artificial immune system for parameter estimation of induction motor. *Expert Systems with Applications* 37.3, 6109-6115.
[11] Sakthivel, V. P., R. Bhuvaneswari & S. Subramanian (2010). Bacterial foraging technique based parameter estimation of induction motor from manufacturer data. *Electric Power Components and Systems.* 38.6, 657-674.
[12] Perez, I., M. Gomez-Gonzalez & F. Jurado (2013). Estimation of induction motor parameters using shuffled frog-leaping algorithm. *Electr Eng.* 95.267-275.
[13] Perez, I., M. Gomez-Gonzalez & F. Jurado (2012). Determination of induction motor parameters with differential evolution algorithm. *Neural Computing & Applications.* 21, 1995-2004.
[14] Gomez-Gonzalez, M. & F. Jurado (2012). Shuffled frog-leaping algorithm for parameter estimation of double cage asynchronous machine. *IET Electric Power Applications* 6.3, 484-490.
[15] Karaboga, D. (2005). An idea based on honey bee swarm for numerical optimization. Technical report TR06, Turkey: Computer Engineering Department University.
[16] Ayan, K. & U. Kilic (2012). Artificial bee colony algorithm solution for optimal reactive power flow. *Appliedl Soft Computing.* 12.5, 1477-1482.
[17] Sahin, A. S., B. Kilic & U. Kilic (2011). Design and economic optimization of shell and tube heat exchangers using artificial bee colony (ABC) algorithm. *Energy Conversion and Management.* 52.11, 3356-3362.
[18] Mukherjr, P. & L. Satish (2012). Construction of equivalent circuit of a single and isolated transformer winding from FRA data using the ABC algorithm. *IEEE Transaction on Power Delivery* 27.2, 963-970.
[19] Karaboga, D. & B. Basturk (2008). On the performance of artificial bee colony (ABC) algorithm. *Applied Soft Computing.* 8.1, 687-697.
[20] Akay, B. & D. Karaboga (2012). A modified artificial bee colony algorithm for real-parameter optimization. *Information Sciences* 192, 120-142.
[21] The Mathworks Inc.: MATLAB Release 2011a, available at http://www.mathworks.com (accessed August 2011).